\documentclass[aps,prb,reprint,amsmath,amssymb,superscriptaddress]
{revtex4-1}
\usepackage{graphicx}

\begin{document}

\title{Anomalous pressure dependence of magnetic ordering temperature in Tb to 141 GPa:  comparison with Gd and Dy}

\author{J.~Lim}
\affiliation {Department of Physics, Washington University, St. Louis, MO 
63130, USA}

\author{G.~Fabbris}
\altaffiliation[Current address: ]{Department of Condensed Matter Physics and Material Science, Brookhaven National Laboratory, Upton, NY 11973, USA }
\affiliation {Department of Physics, Washington University, St. Louis, MO 63130, USA}
\affiliation {Advanced Photon Source, Argonne National Laboratory, Argonne, IL 
60439, USA}

\author{D.~Haskel}
\affiliation {Advanced Photon Source, Argonne National Laboratory, Argonne, IL 
60439, USA}

\author{J.~S.~Schilling}
\email[]{jss@wuphys.wustl.edu}
\affiliation {Department of Physics, Washington University, St. Louis, MO 63130, USA}

\date{\today}

\begin{abstract}
In previous studies the pressure dependence of the magnetic ordering
temperature $T_{\text{o}}$ of Dy was found to exhibit a sharp increase above
its volume collapse pressure of 73 GPa, appearing to reach temperatures well
above ambient at 157 GPa. In a search for a second such lanthanide,
electrical resistivity measurements were carried out on neighboring Tb to
141 GPa over the temperature range 3.8 - 295 K. Below Tb's volume collapse
pressure of 53 GPa, the pressure dependence $T_{\text{o}}(P)$ mirrors that
of both Dy and Gd. However, at higher pressures $T_{\text{o}}(P)$ for Tb
becomes highly anomalous. This result, together with the very strong
suppression of superconductivity by dilute Tb ions in Y, suggests that
extreme pressure transports Tb into an unconventional magnetic state with an
anomalously high magnetic ordering temperature.
\end{abstract}

\maketitle

\section{Introduction}

The magnetic ordering temperatures $T_{\text{o}}$ of Gd and Dy have been
recently shown to track each other in a highly non-monotonic fashion as a
function of pressure to $\sim 70$ GPa; at higher pressures they deviate
markedly, $T_{\text{o}}$ for Dy rising rapidly to temperatures well above
ambient at 157 GPa.\cite{lim0} Parallel experiments on dilute magnetic
alloys of Gd and Dy with superconducting Y suggest that for pressures above $%
\sim 70$ \ GPa Dy is transformed from a magnetically conventional lanthanide
into one with an unconventional magnetic state with marked deviations from
de Gennes scaling,\cite{degennes} a state perhaps governed by Kondo physics,
indicating that the Dy ion is nearing a magnetic instability.\cite{lim0} An
alternate explanation is that the strong enhancement of $T_{\text{o}}$ in Dy
arises through changes in the crystalline electric field at extreme pressure.%
\cite{lim0} Analogous studies on additional lanthanides are recommended to
help identify the origin of this anomalous behavior.

The lanthanide Tb, which lies between Gd and Dy in the periodic table, has
one fewer 4$f$ electron than Dy, and is probably less stable magnetically
than Dy due to its direct proximity to Gd, by far the most stable of all
magnetic lanthanides. Tb orders antiferromagnetically (AFM) at $T_{\text{o}%
}\simeq $ 230 K followed by a ferromagnetic (FM) transition at $T_{\text{o}%
}\simeq $ 220 K.\cite{koehler1} Both transition temperatures initially
decrease rapidly with pressure at the rate -10 to -12 K/GPa, but above $\sim
7$ GPa neither transition can be clearly detected in either the ac or dc
magnetic susceptibility.\cite{mcwhan,jackson1,mito} The disappearance of the
ordered moment in the susceptibility measurement indicates a transition to
either an AFM or paramagnetic state above 7 GPa. Electrical resistivity
studies should reveal which scenario is correct since both FM and AFM order
normally lead to a distinct kink in the temperature dependence of the resistivity.
However, recent resistivity and neutron diffraction experiments on Tb find that the FM transition decreases with pressure at the rate -16.7 K/GPa to 3.6 GPa;\cite{thomas} that the transition could no longer be resolved above 3.6 GPa may be due to appreciable pressure-gradient broadening in the cell which contained no pressure medium. That magnetic order in Tb disappears above 7 GPa seems highly unlikely since both
x-ray absorption near-edge structure (XANES) and non-resonant x-ray emission
spectroscopy (XES) measurements detect no change in Tb's valence to 65 GPa
and 70 GPa, respectively.\cite{fabbris1} In fact, the XES studies show that
Tb retains its strong, highly localized magnetic moment ($J=6$) to at least
70 GPa.\cite{fabbris1}

In this paper we present the results of dc electrical resistivity
measurements on Tb over the temperature range 3.8 - 295 K to pressures as
high as 141 GPa, well above the pressure of 53 GPa where Tb suffers a 5\%
volume collapse at the phase transition from hexagonal hR24 to body-centered
monoclinic (bcm).\cite{cunningham} Magnetic order is indeed observed in Tb
for pressures above 7 GPa. In fact, to 53 GPa $T_{\text{o}}(P)$ follows nearly the
same highly non-monotonic pressure dependence found earlier in Gd and Dy,\cite{lim0} but
deviates markedly at higher pressures. As the applied
pressure passes through 53 GPa, $T_{\text{o}}(P)$ for Tb first decreases,
but then begins to increase rapidly above 80 GPa. As suggested for Dy,\cite%
{lim0} extreme pressure appears to transport Tb into an unconventional
magnetic state with an anomalously high magnetic ordering temperature, well
above that anticipated from conventional de Gennes scaling.

\section{Experimental Techniques}

Resistivity samples were cut from a Tb ingot (99.9\% pure, Material
Preparation Center of the Ames Laboratory\cite{ames}). To generate pressures
well beyond the volume collapse pressure of Tb at 53 GPa, a diamond anvil
cell (DAC) made of CuBe alloy was used.\cite{Schilling84} Two separate
high-pressure experiments were carried out where pressure was generated by
two opposing diamond anvils (1/6-carat, type Ia) with 0.35 mm diameter
culets beveled at 7$^{\circ }$ to 0.18 mm central flats.

The Re gasket (6 - 7 mm diameter, 250 $\mu $m thick) was preindented to 30 $%
\mu $m and a 80 $\mu $m diameter hole electro-spark drilled through the
center. The center section of the preindented gasket surface was filled with
a 4:1 cBN-epoxy mixture to insulate the gasket and serve as pressure medium.
The thin Tb sample was then placed on top of four thin (4 $\mu $m) Pt leads
for a four-point dc electrical resistivity measurement. In an attempt to
minimize the effect of the pressure gradient across the sample in this
non-hydrostatic pressure environment, in run 1 an elongated sample
(dimensions $\sim $8$\times $80$\times $3 $\mu $m$^{3}$) was used with the
two voltage leads spaced only 5 $\mu $m apart (see inset to Fig. 1(a)). In
run 2 all four Pt leads were placed near the corners of the square-shaped
sample (dimensions $\sim $30$\times $30$\times $5 $\mu $m) (see inset to
Fig. 1(b)), as in the previous resistivity measurements on Dy.\cite{lim0}
However, from the temperature-dependent resistivity data the pressure
gradient was estimated to be approximately the same in both runs. Further
details of the non-hydrostatic high pressure resistivity technique are given
in a paper by Shimizu \textit{et al}.\cite{Shimizu05}

\begin{figure*}[t]
\includegraphics[width = 14 cm]{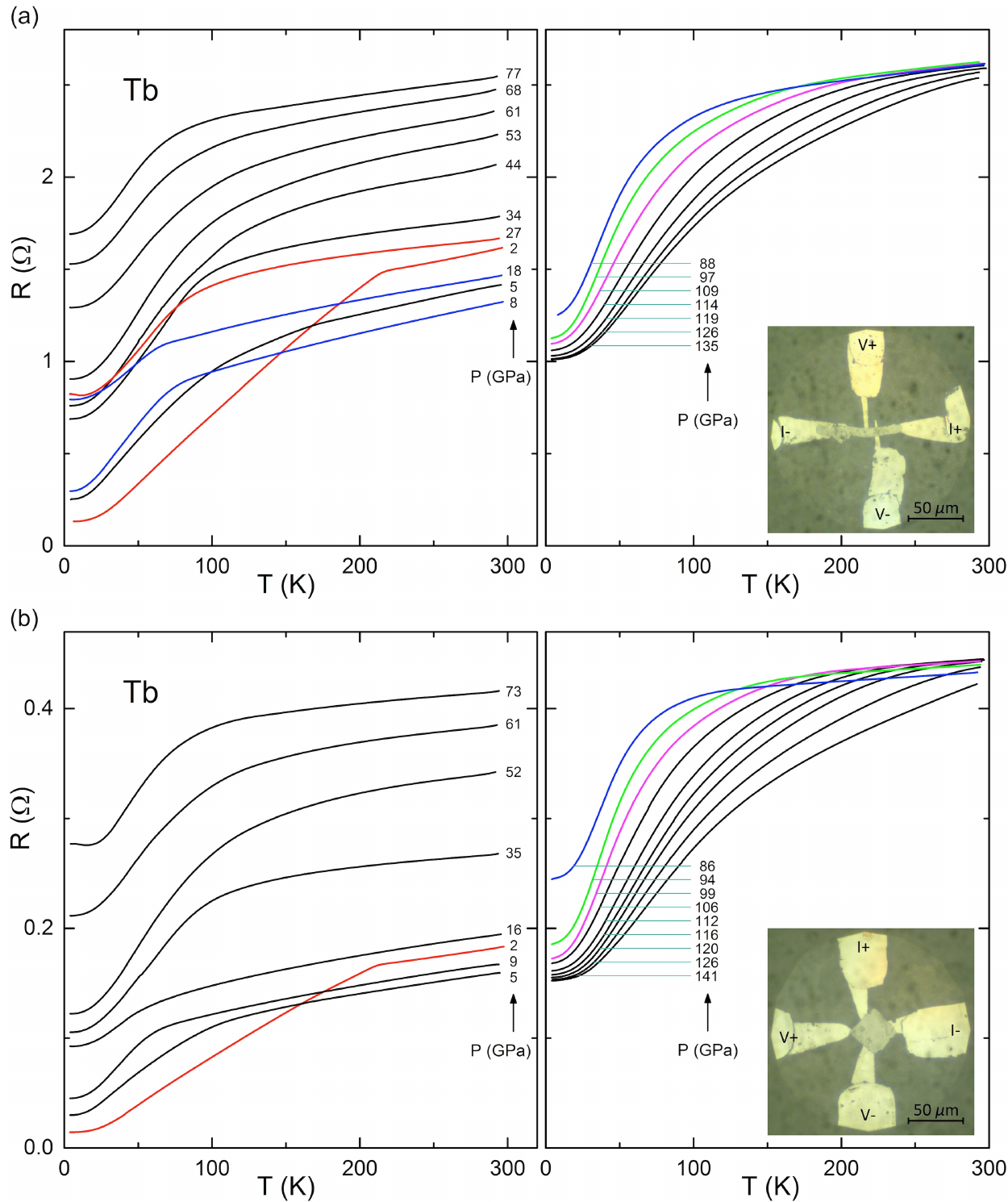} 
\caption{\label{fig1}(color online) Resistance of Tb versus temperature to
295 K for (a) run 1 and (b) run 2 at various pressures. Insets show
photograph of (a) elongated sample in run 1 and (b) square sample in run 2.}
\end{figure*}

A He-gas driven membrane was utilized to change pressure at any temperature.%
\cite{daniels1} The value of the pressure was determined using both the
fluorescence\cite{Chijioke05} from a small ruby sphere positioned at the
edge of the sample and the frequency shift of the diamond vibron via Raman
spectroscopy.\cite{raman1} The ruby pressure was determined at both ambient
temperature and a temperature within 20 K of $T_{\text{o}}$; the vibron
pressure was determined only at ambient temperature. The values of the
pressure given are averaged over the sample to an estimated accuracy of $\pm
10\%.$ In these experiments temperatures from 3.8 K to 295 K were available
using an Oxford flow cryostat. All measurements shown in this paper were
carried out with increasing pressure; diamond anvil failure at the highest
pressure ended the experiment. Further experimental details of the DAC and
cryostat are given elsewhere.\cite{fabbris1,Schilling84,klotz1,debessai1}

\section{Results of Experiment}

The present resistivity studies on Tb were carried out in two separate
experiments. In Fig. 1(a) the electrical resistance $R(T)$ from run 1 is
plotted versus temperature at 18 different pressures to 135 GPa. The results
from run 2 are shown in Fig. 1(b) and span the pressure range 2 - 141 GPa
with 17 values. The onset of magnetic ordering is identified by the kink in
the $R(T)$ dependence clearly seen near 200 K at 2 GPa, the lowest pressure
in each run. The kink in $R(T)$ upon cooling marks the beginning of the
suppression of spin-disorder scattering $R_{sd}(T)$ as magnetic ordering
sets in.\cite{ref3} At higher pressures this kink broadens somewhat into a
"knee" due to an increasing pressure gradient across the sample, but remains
clearly visible to $\sim $115 GPa.

In Fig. 2 selected data from Fig. 1(b) are replotted but shifted vertically
for clarity so that no curves intersect. In this graph the red line through
the data above the knee gives the phonon contribution $R_{ph}(T)$ to the
total measured resistance $R(T)$ estimated in the same manner as in our
previous work on Dy,\cite{lim0} as outlined in the next paragraph. The
paramagnetic state of Tb yields the relatively flat region of $R(T)$ at
higher temperature where the red (phonon) line overlaps the data. Where the
red line begins to separate from the data marks the onset of magnetic
ordering in some region of the sample. Because of the pressure gradient
across the sample, other regions of the sample will have a lower onset
temperature, thus broadening the kink into a knee. The intersection of the
phonon resistance (red curve) with the red low-temperature tangent curve
defines the temperature $T_{\text{x}}$ in Fig. 2.

\begin{figure}[t]
\includegraphics[width = 8.5 cm]{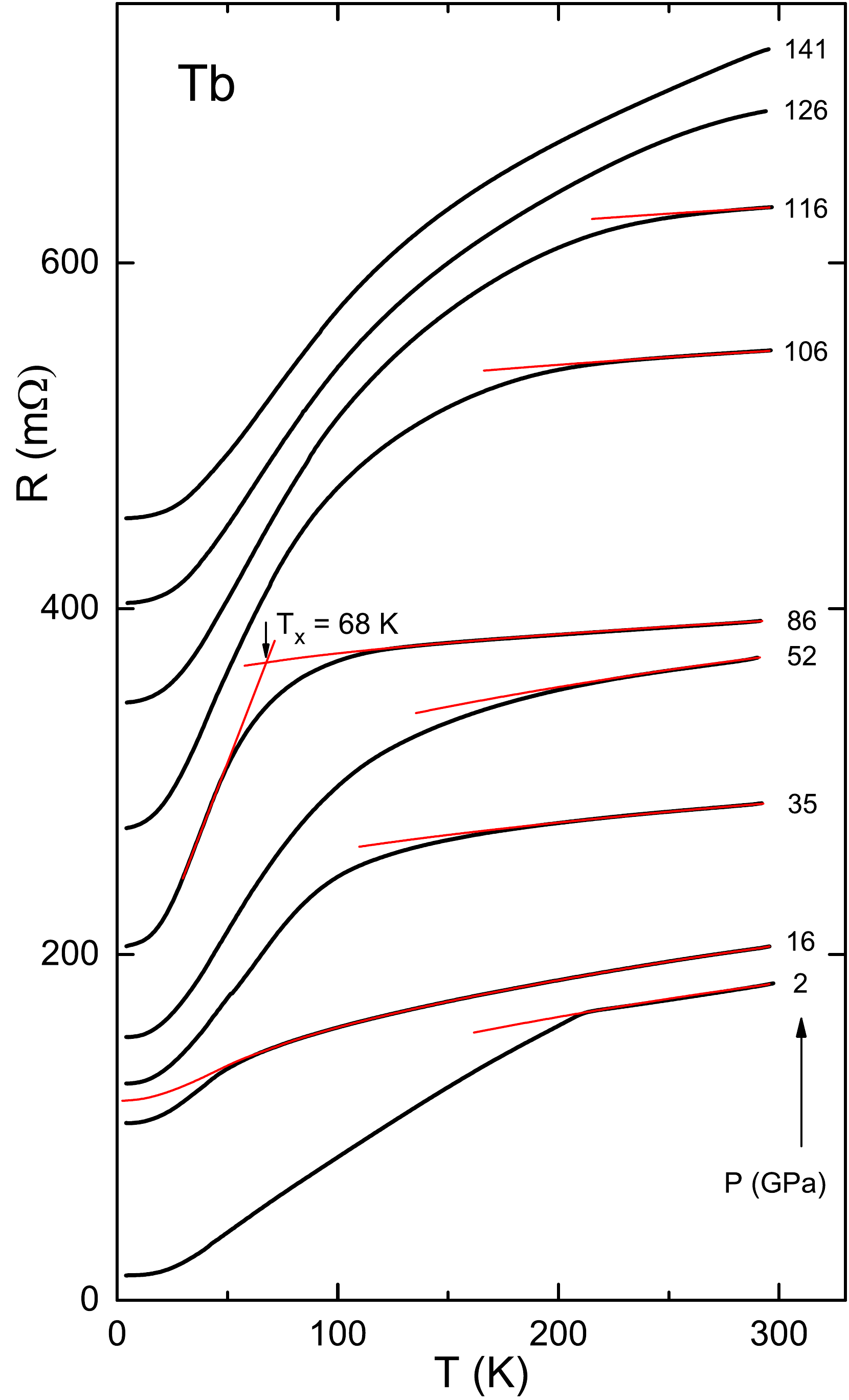} 
\caption{\label{fig2}(color online) Selection of resistance versus
temperature curves for Tb from run 2 in Fig. 1(b) where, except at 2 GPa,
the curves have been shifted vertically for clarity. Red lines with small
positive slope give temperature dependence of phonon resistance for $%
T\gtrsim T_{\text{x }}$except at 16 GPa where the phonon resistance extends
to 0 K (see text).}
\end{figure}

The total measured resistance is the sum of three terms, $%
R(T)=R_{d}+R_{ph}(T)+R_{sd}(T),$ where $R_{d}$ is the
temperature-independent defect resistance, $R_{ph}(T)$ the
temperature-dependent phonon resistance, and $R_{sd}(T)$ the
temperature-dependent spin-disorder resistance. At temperatures where there
is no magnetic ordering in the sample, $R_{sd}(T)$ is independent of
temperature. Above the onset temperature of the knee, the temperature
dependence of $R(T)$ is, therefore, due solely to that of $R_{ph}(T).$ The
temperature dependence of the phonon resistance is visible over the widest
temperature range at that pressure (16 GPa in Fig. 2) where the knee begins
at the lowest temperature. We extrapolate this dependence to 0 K in the
temperature region below the knee to yield the temperature-dependent
function $R_{ph}^{16}(T)$, the estimated phonon resistance at 16 GPa in run
2. In run 1 the data at 18 GPa were used in the same way to obtain $R_{ph}^{18}(T)$. Since the functional dependence of $R_{ph}(T)$ on temperatures above $T_{%
\text{x}}$ is seen in Fig. 2 to change only slowly with pressure, we
estimate $R_{ph}(T)$ for the other pressures in run 2 by simply multiplying the
function $R_{ph}^{16}(T)$ by a "phonon factor" $\alpha $ chosen such that
for temperatures above the knee the quantity $R(T)-\alpha R_{ph}^{16}(T)$
becomes \textit{temperature independent} for $T>T_{\text{x}}$. The values of 
$\alpha $ required are listed in Table 1 at all pressures in run 2 to 141
GPa. For pressures of 120 GPa and above, the knee in $R(T)$ apparently
begins above 295 K, so that $\alpha $ can no longer be estimated directly
from the resistance data. For $P\geq 120$ GPa, therefore, the value $\alpha
=0.41$ is assumed in Table 1 for run 2 and $\alpha
=0.69$ in run 1. This is admittedly an oversimplified way to
estimate the phonon contribution, but is superior to the assumption made in
an earlier study that for many lanthanides $R_{ph}$ is simply a linear
function of temperature to 0 K.\cite{colvin1}

In Fig. 3 the extracted spin-disorder resistance $R_{sd}(T)=R(T)-\alpha
R_{ph}^{16}(T)-R_{d}$ is plotted for pressures 5, 35, and 86 GPa in run 2.
The saturation (maximum) value of the so obtained spin-disorder resistance $%
R_{sd}^{\max }$ in the paramagnetic phase at each pressure is listed in
Table 1 for run 2. At 86 GPa, for example, $%
R_{sd}^{\max }\simeq 152$ m$\Omega $ as seen in Fig. 3. A similar procedure was used to obtain $R_{sd}^{\max }$ in run 1.

\begin{figure}[t]
\includegraphics[width = 8.5 cm]{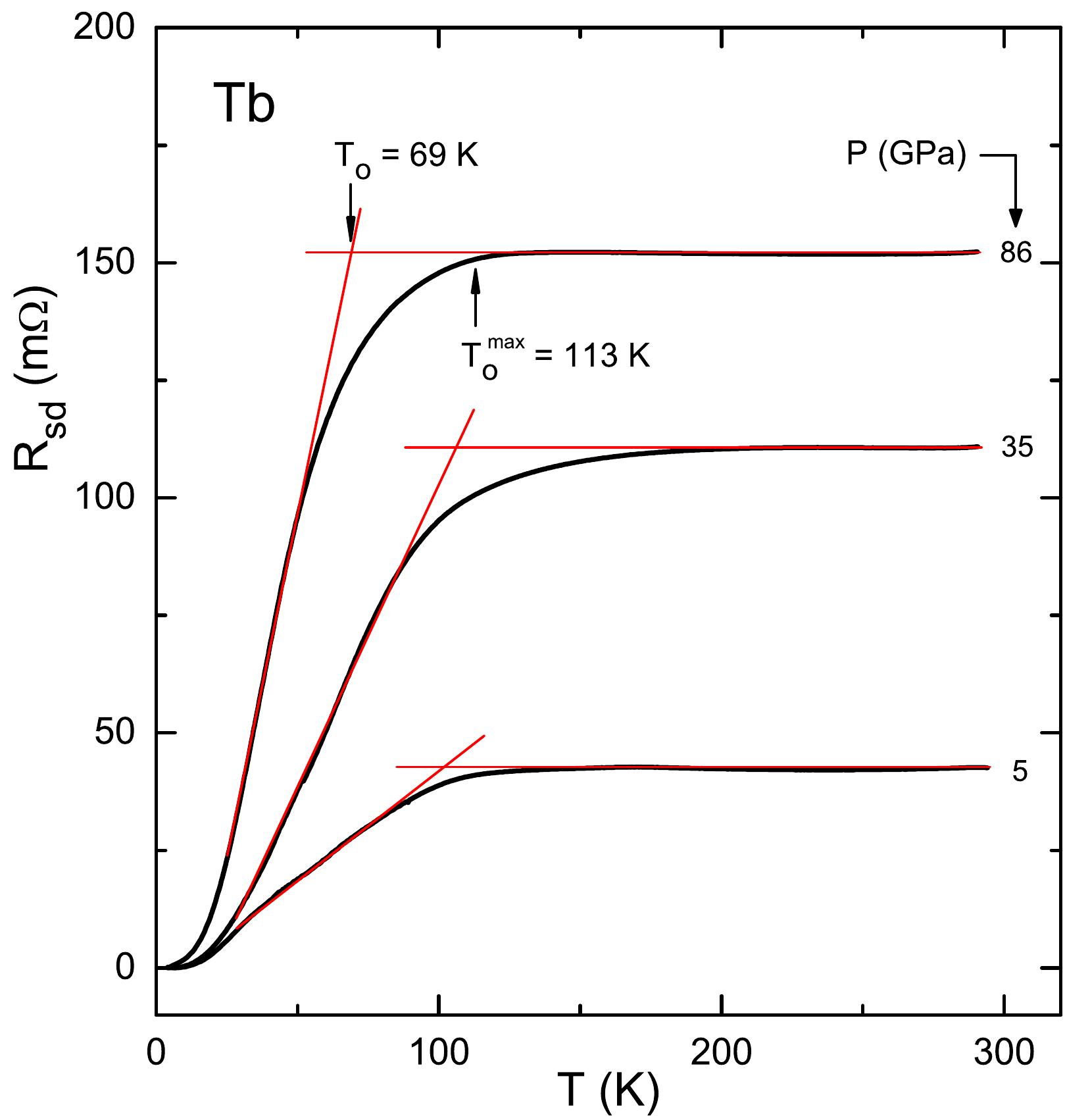} 
\caption{\label{fig3}(color online) Spin-disorder resistance $R_{sd}(T)$
versus temperature at three pressures from run 2. The phonon $R_{ph}(T)$ and
defect $R_{d}$ resistances have been subtracted off. The average magnetic
ordering temperature $T_{\text{o}}$ is defined by intersection point of two
tangent lines. $T_{\text{o}}^{\max }$ gives temperature at which
spin-disorder resistance has decreased by 1\% (see text).}
\end{figure}

As illustrated in Fig. 3, the average magnetic ordering temperature $T_{%
\text{o}}$ in the Tb sample is estimated from the point of intersection of
two straight (red) lines, a horizontal line for temperatures above the onset
of the knee, and a line tangent to $R_{sd}(T)$ at lower temperatures. For 86
GPa it is seen that $T_{\text{o}}\simeq 69$ K. This temperature differs by
only 1 K from $T_{\text{x}}\simeq 68$ K, the intersection point of the
phonon resistance and the low-temperature tangent lines in Fig. 2 at the
same pressure. Here we regard $T_{\text{o}}$ to be the average magnetic
ordering temperature (in our previous paper on Dy, $T_{\text{x}}$ was used
as the ordering temperature\cite{lim0}). Since the pressure gradient leads
to a variation in the value of the magnetic ordering temperature across the
sample, we define the \textquotedblleft maximum\textquotedblright  ordering temperature $T_{\text{o}}^{\max }$
as the temperature at which the spin-disorder resistance has decreased by
1\%. In Fig. 3 it is seen that $T_{\text{o}}^{\max }\simeq 113$ K at 86 GPa.
If $dT_{\text{o}}/dP>0,$ $T_{\text{o}}^{\max }$ gives the value of the
magnetic ordering temperature at the center of the cell (sample) where the
pressure is highest. In Fig. 3 it is seen that $T_{\text{o}}^{\max }$ lies
44 K higher than $T_{\text{o}}$ at 86 GPa. All values of $T_{\text{o}}$ and $%
T_{\text{o}}^{\max }$ in run 2 are listed in Table 1.

In Fig. 4 $T_{\text{o}}$ and $T_{\text{o}}^{\max }$ are plotted versus
pressure to 141 GPa for runs 1 and 2 on Tb; values for $P\gtrsim 120$ GPa
are estimated using a procedure from Ref. \onlinecite{lim0}, as outlined below.
Where they can be compared, the present results are in reasonable agreement
with earlier ac magnetic susceptibility measurements of Jackson \textit{et
al.} to 6.3 GPa.\cite{jackson1} The pressure dependence $T_{\text{o}}(P)$ at
higher pressures is seen to be highly non-monotonic, presumably in response
to multiple structural phase transitions\cite{cunningham} (see top of the
graph). Note that the phase boundaries were determined from x-ray
diffraction studies at ambient temperature and may shift somewhat as the
temperature is lowered.

\begin{figure}[t]
\includegraphics[width = 8.5 cm]{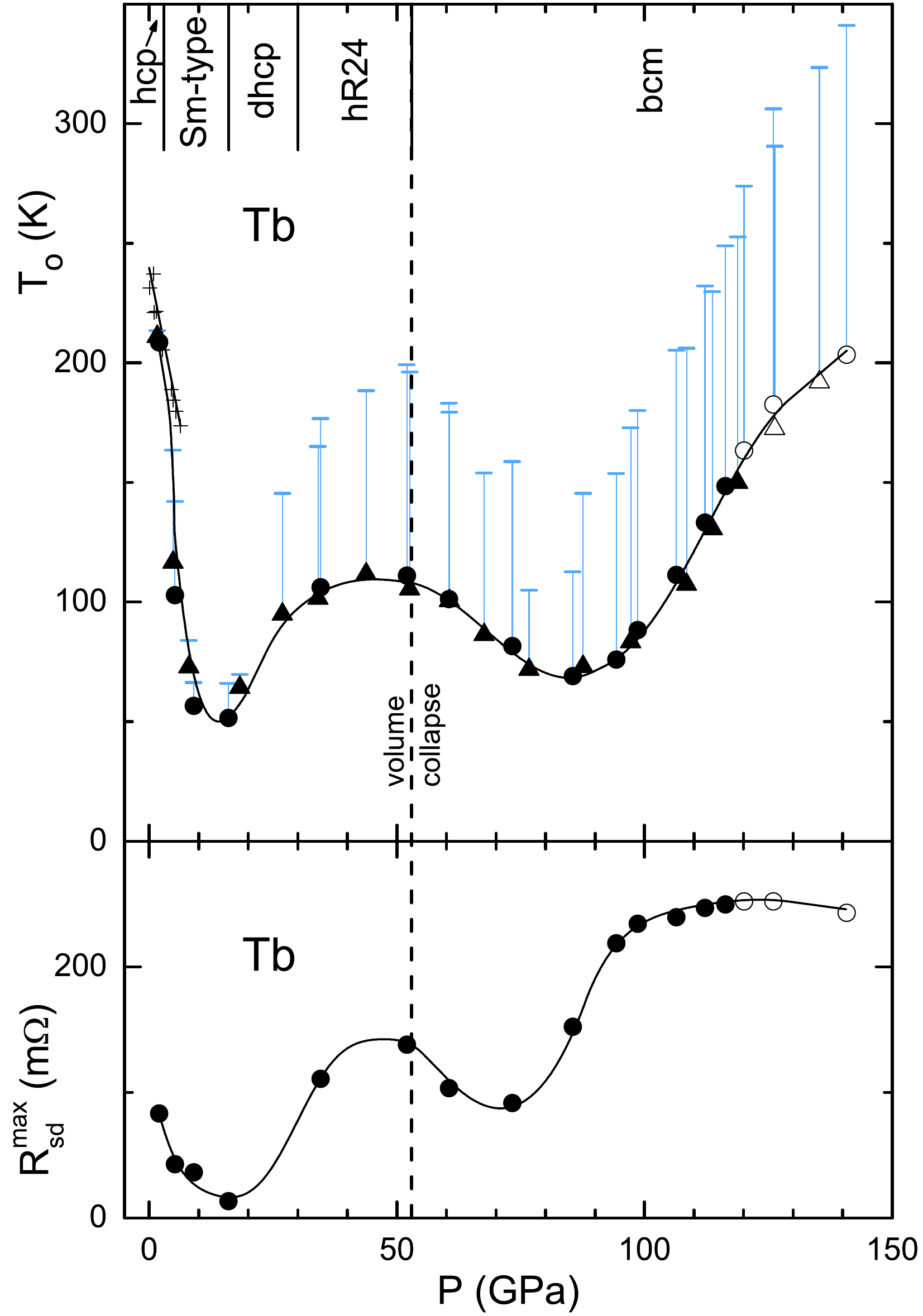} 
\caption{\label{fig4}(color online) Average magnetic ordering temperatures $T_{\text{o}}$ of
Tb versus pressure: ($+$) earlier susceptibility studies to 6.3 GPa with
slope $dT_{\text{c}}/dP=-11$ K/GPa;\cite{jackson1} present resistance
measurements to 141 GPa from ($\blacktriangle $) run 1, ($\bullet $) run 2.
\textquotedblleft error bar\textquotedblright  connected vertically to each value of $T_{\text{o}}$ gives
maximum ordering temperature $T_{\text{o}}^{\max }$\ at that pressure. Open
symbols indicate extrapolated values (see text). Vertical dashed line marks
pressure of volume collapse for Tb at 53 GPa. Crystal structures for Tb are
given at top of graph.\cite{cunningham} $R_{sd}^{\max }$ versus
pressure is plotted in lower part of figure from run 2 where it is seen to roughly track $T_{\text{o}}$(P).}
\end{figure}

A comparison of $T_{\text{o}}(P)$ for Tb from Fig. 4 to comparable graphs
for Gd and Dy in Ref. \onlinecite{lim0} reveals a remarkable similarity to 53 GPa,
the pressure at which the 5\% volume collapse in Tb occurs.\cite{cunningham}
Also plotted in Fig. 4 are the values of $T_{\text{o}}^{\max }$ for Tb given
by the upper error bars connected to the values of $T_{\text{o}}$ at each
pressure by a light (blue) vertical line. Particularly intriguing is the
decrease in $T_{\text{o}}$ following the hR24 to body-centered monoclinic
(bcm) transition at 53 GPa,\cite{cunningham} followed by a rapid increase
above 80 GPa. In contrast to the findings for $P\leq 53$ GPa, at higher
pressures $T_{\text{o}}(P)$ for Tb thus differs significantly from that
found earlier for either Gd or Dy.\cite{lim0} Plotted versus relative volume 
$V/V_{\text{o}}$, the increase of $T_{\text{o}}$ above 80 GPa for Tb is found to be much
more rapid than the initial decrease of  $T_{\text{o}}$ to 6.3 GPa. A
similar result was found for Dy.\cite{lim0} Extrapolating $T_{\text{o }}$
versus $V/V_{\text{o }}$ for Tb linearly to $V/V_{\text{o}}=0.40$ (141 GPa),
yields the values $T_{\text{o}}\approx $ 250 K and $T_{\text{o}}^{\max
}\approx 350$ K.

We now attempt a more quantitative estimate of the pressure dependence of $%
T_{\text{o}}$, $T_{\text{o}}^{\max }$, and $R_{sd}^{\max }$ in the pressure
range above $116$ GPa where the onset of the knee appears to lie at or above
ambient temperature. We first consider the spin-disorder resistance $%
R_{sd}(T)$ at pressures $P<120$ GPa. The first step is to normalize $%
R_{sd}(T)$ to its value at 295 K, yielding the relative spin-disorder
resistance $R_{sd}(T)/R_{sd}^{\max }$ plotted versus $\log T$ for data at
106, 112, and 116 GPa in Fig. 5. Since at the higher pressures of 120, 126,
and 141 GPa the onset of magnetic ordering appears to lie above the
temperature range of the present experiments (295 K), one cannot determine
the value of $R_{sd}$ in the paramagnetic phase, nor $T_{\text{o}}$ or $T_{%
\text{o}}^{\max },$ directly from the resistance data. However, noticing
that over much of the temperature range the $R_{sd}(T)$ curves for 106, 112,
and 116 GPa are approximately parallel on the $\log T$ plot in Fig. 5, we
divide the $R_{sd}(T)$ data for $P\geq 120$ GPa by that factor which results
in curves parallel to those at the lower pressures, as seen in Fig. 5. We
identify this factor as the value of the temperature-independent
spin-disorder resistance $R_{sd}^{\max }$ in the paramagnetic phase, as
listed in Table 1. This procedure is tantamount to assuming that $%
R_{sd}=R_{sd}(T/T_{\text{o}})$ for $P\geq 106$ GPa. The change in the value
of the magnetic ordering temperatures $T_{\text{o}}$ and $T_{\text{o}}^{\max
}$ can now be estimated from the shift of the $R_{sd}(T)/R_{sd}^{\max }$
curves along the $\log T$ axis. The resulting values of $T_{\text{o}}(P)$
and $T_{\text{o}}^{\max }(P)$ are given in Table 1 and in Fig. 4 for all
pressures in runs 1 and 2 as the open triangles and circles, respectively. From this analysis we infer that from 116 to 141
GPa the average magnetic ordering temperature $T_{\text{o}}$ has increased
from 148 K to 203 K and the maximum ordering temperature $T_{\text{o}}^{\max
}$ from 249 K to 341 K, values close to those obtained in the linear
extrapolation above. 

\begin{figure}[t]
\includegraphics[width = 8.5 cm]{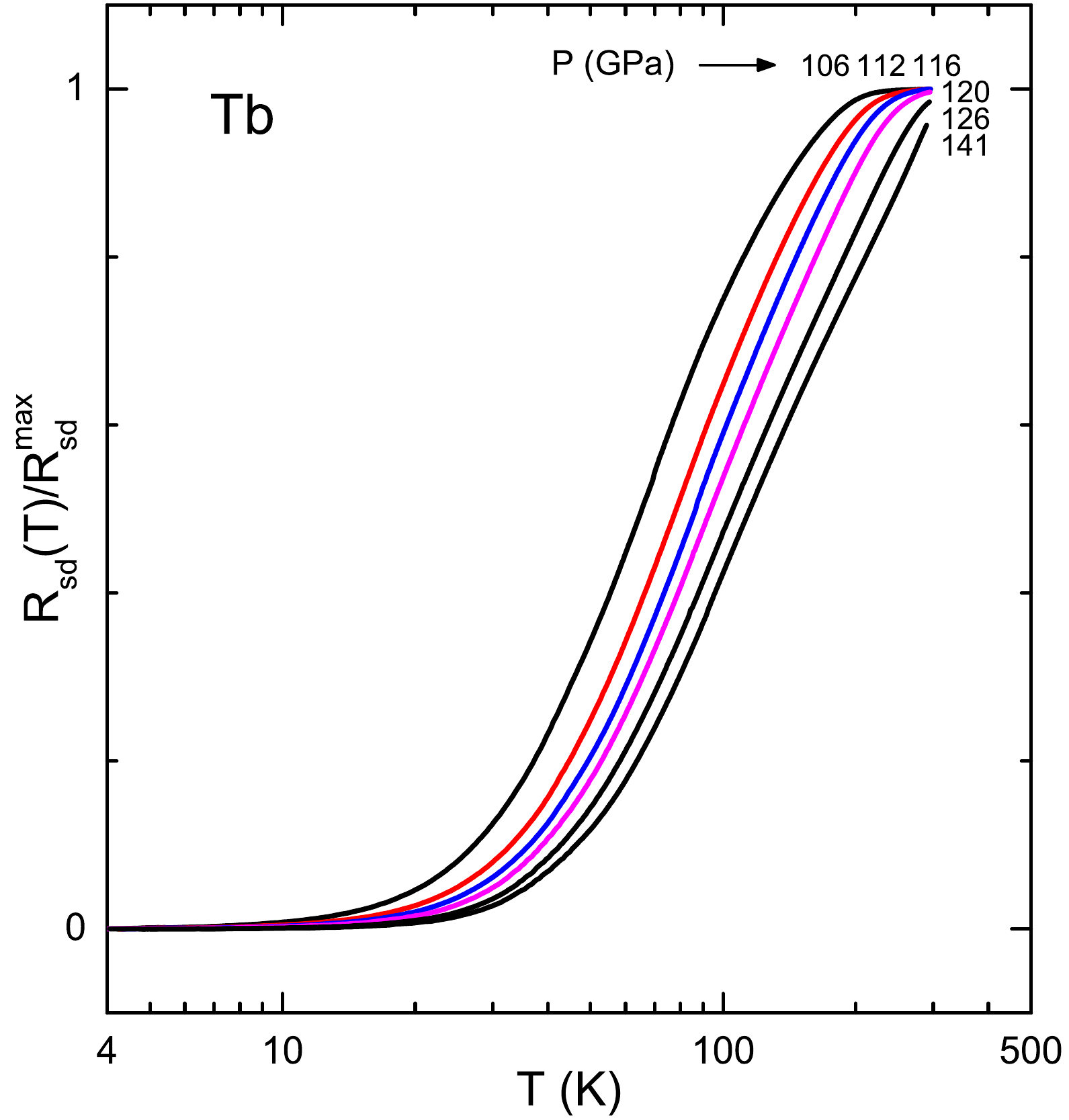} 
\caption{\label{fig5}(color online) Relative spin-disorder resistance $%
R_{sd}(T)/R_{sd}^{\max }$ versus $\log T.$ For data at 120, 126, and 141
GPa, $R_{sd}^{\max }$ is estimated by adjusting slope of temperature
dependence to match that at 106, 112 and 116 GPa (see text). From relative
horizontal shifts of the curves the pressure-dependence of the magnetic
ordering temperature $T_{\text{o}}$ is estimated for 120, 126, and 141 GPa
(see text and Table 1).}
\end{figure}

In our previous work on Dy the spin-disorder resistance in the paramagnetic
state $R_{sd}^{\max }$ was found to approximately track the magnetic
ordering temperature $T_{\text{o}}$ as a function of pressure. This same
result is seen in Fig. 4 to hold for Tb.

\section{Discussion}

We now seek to identify the mechanism(s) responsible for the highly
non-monotonic dependence of Tb's magnetic ordering temperature $T_{\text{o}}$
on pressure. First we focus on the pressure region below $53$ GPa, the
pressure at which Tb suffers a 5\% volume collapse. Since the pressure
dependence of $T_{\text{o }}$ is so similar for Tb, Gd and Dy in this
pressure range, a common mechanism seems likely.

For a conventional lanthanide metal with a stable magnetic moment, the
magnetic ordering temperature $T_{\text{o }}$ is expected to scale with the
de Gennes factor $(g-1)^{2}J_{t}(J_{t}+1)$, modulated by the prefactor $%
J^{2}N(E_{\text{F}})$, where $J$ is the exchange interaction between the 4$f$
ion and the conduction electrons, $N(E_{\text{F}})$ the density of states at
the Fermi energy, $g$ the Land\'{e}-$g$ factor, and $J_{t}$ the total
angular momentum quantum number.\cite{degennes} Since the de Gennes factor
is constant under pressure, unless the magnetic state becomes unstable
and/or a valence transition occurs, the marked similarity between the highly
non-monotonic pressure dependences of $T_{\text{o}}$ for Tb, Dy and Gd to 53
GPa likely originates in the pressure dependence of the prefactor $J^{2}N(E_{%
\text{F}})$, facilitated by a series of nearly identical structural phase
transitions in Tb,\cite{cunningham} Dy,\cite{patterson1} and Gd.\cite%
{gd,erran} These phase transitions are likely driven by increasing 5\textit{d%
}-electron occupation with pressure.\cite{pettifor} Indeed, electronic
structure calculations for Dy suggest that its large negative initial
pressure derivative $dT_{\text{o}}/dP\simeq -6.7$ K/GPa results from a
strong decrease in $J^{2}N(E_{\text{F}})$.\cite{jackson1,fleming}

\begin{table}[t]
\caption{\label{table1} Values for Tb of the average $T_{%
\text{o}}$ and maximum $T_{\text{o}}^{\max }$ magnetic ordering
temperatures, spin-disorder resistance $R_{sd}^{\max }$ for $T>T_{\text{o}%
}^{\max }$, and phonon factor $\alpha $ as a function of pressure from runs 1 and 2 (see text).}
\begin{ruledtabular}
\begin{tabular}{c c c c c c}
run & $P$(GPa) & $T_{\text{o}}$(K) & $T_{\text{o}}^{\max }$(K) & $R_{sd}^{\max }($m%
$\Omega )$ & $\alpha $ \\ 
\hline
1 & 2 & 211 & 214 & 881 & 1.16 \\
1 & 5 & 117 & 163 & 523 & 1.24 \\ 
1 & 8 & 73 & 84 & 380 & 1.24 \\ 
1 & 18 & 64 & 70 & 154 & 1.0 \\ 
1 & 27 & 95 & 145 & 513 & 0.66 \\ 
1 & 34 & 101 & 165 & 721 & 0.72 \\ 
1 & 44 & 112 & 188 & 769 & 1.02 \\ 
1 & 53 & 105 & 196 & 746 & 1.11 \\ 
1 & 61 & 101 & 179 & 518 & 1.05 \\ 
1 & 68 & 86 & 154 & 479 & 0.89 \\ 
1 & 77 & 72 & 105 & 469 & 0.74 \\ 
1 & 88 & 73 & 145 & 1001 & 0.69 \\ 
1 & 97 & 83 & 173 & 1140 & 0.69 \\ 
1 & 109 & 107 & 206 & 1165 & 0.69 \\ 
1 & 114 & 131 & 230 & 1189 & 0.69 \\ 
1 & 119 & 150 & 253 & 1201 & 0.69 \\ 
1 & 126 & 172 & 291 & 1211 & 0.69 \\ 
1 & 135 & 192 & 323 & 1214 & 0.69 \\ 
\hline
2 & 2 & 208 & 210 & 83 & 0.98 \\ 
2 & 5 & 103 & 142 & 43 & 0.98 \\ 
2 & 9 & 57 & 66 & 36 & 0.98 \\ 
2 & 16 & 52 & 66 & 13 & 1.0 \\ 
2 & 35 & 106 & 177 & 111 & 0.58 \\ 
2 & 52 & 111 & 199 & 138 & 0.92 \\ 
2 & 61 & 101 & 183 & 103 & 0.79 \\ 
2 & 73 & 82 & 159 & 91 & 0.56 \\ 
2 & 86 & 69 & 113 & 152 & 0.41 \\ 
2 & 94 & 76 & 154 & 219 & 0.41 \\ 
2 & 99 & 88 & 180 & 234 & 0.41 \\ 
2 & 106 & 111 & 205 & 239 & 0.41 \\ 
2 & 112 & 133 & 232 & 247 & 0.41 \\ 
2 & 116 & 148 & 249 & 250 & 0.41 \\ 
2 & 120 & 163 & 274 & 252 & 0.41 \\ 
2 & 126 & 183 & 306 & 252 & 0.41 \\ 
2 & 141 & 203 & 341 & 243 & 0.41 \\ 
\end{tabular}
\end{ruledtabular}
\end{table}

We now consider the pressure region $P>$ 53 GPa where the pressure
dependence $T_{\text{o }}(P)$ for Tb is highly anomalous, deviating markedly
from that of the model conventional lanthanide Gd to at least 127 GPa.\cite%
{lim0} The absence of magnetic instabilities in Gd, even at extreme
pressures, is expected since the local magnetic state of Gd with its
half-filled 4$f^{7}$ shell is the most stable of all elements, its 4$f^{7}$
level lying $\sim 9$ eV below the Fermi level.\cite{yin} Why is $T_{\text{o}%
}(P)$ in Tb anomalous for $P>$ 53 GPa? A long-standing strategy\cite%
{matthias,maple2} to probe the magnetic state of a given ion is to alloy
this ion in dilute concentration with a host superconductor and determine $%
\Delta T_{\text{c}}$, the degree of suppression of the host's
superconducting transition temperature. Yttrium (Y) is the ideal host
superconductor for Tb since the character of its $spd$-electron conduction
band closely matches that of the heavy lanthanides, Y even exhibiting nearly
the same sequence of structural transitions under pressure.\cite{samu2} One
may thus anticipate that changes in the magnetic state of the Tb ion in the
dilute alloy will be mirrored in the changes occurring in the magnetic state
of Tb metal.

The efficacy of this strategy is supported by previous studies of the
pressure dependences $T_{\text{o}}(P)\ $for Dy metal and $\Delta T_{\text{c}%
}(P)$ for Y(Dy) where both experience a dramatic enhancement beginning just
above the pressure of Dy's volume collapse at $73$ GPa.\cite{lim0} It was
argued that this anomalous behavior might be the result of the Dy ion
exhibiting Kondo physics at elevated pressures where both $T_{\text{o}}$ and 
$\Delta T_{\text{c}}$ are proportional to $\left\vert J_{-}\right\vert ^{2}$%
, the square of the negative exchange parameter leading to the Kondo effect.
Dy's volume collapse itself has been suggested to have its origin in the
Kondo volume collapse model of Allen and Martin.\cite{allen}

\begin{figure}[t]
\includegraphics[width = 8.5 cm]{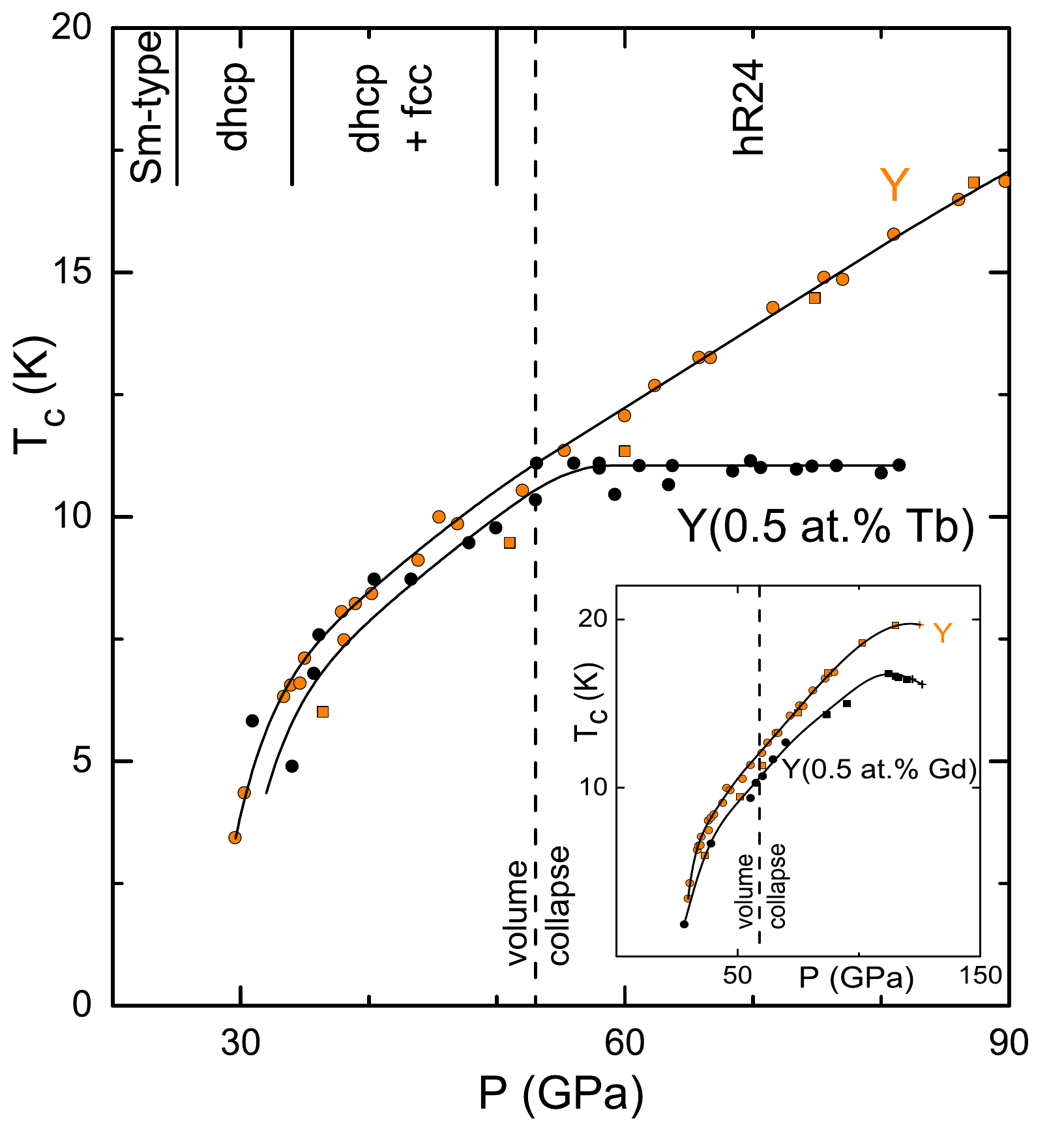} 
\caption{\label{fig6}(color online)  $T_{\text{c}}$ versus pressure for
Y(0.5 at.\% Tb) compared to that for Y, inset showing similar graph for
Y(0.5 at.\% Gd).\cite{fabbris1} Vertical dashed line marks pressure of
volume collapse for Tb at 53 GPa \cite{cunningham} and in inset for Gd at 59
GPa.\cite{gd,erran} At top of graph are crystal structures taken on by superconducting host Y.\cite{samu2}}
\end{figure}

Does perhaps the same scenario apply for Tb? In Fig. 6 the pressure
dependence of the superconducting transition temperature $T_{\text{c}}(P)$
of the dilute magnetic alloy Y(0.5 at.\% Tb) from our previous work\cite%
{fabbris1} is compared to that for elemental Y metal.\cite{hamlin1} To a
pressure of $\sim $50 GPa, $T_{\text{c}}$ for the dilute magnetic alloy is
seen to increase with pressure at the same rate as for Y. However, just
above the pressure of Tb's volume collapse at 53 GPa, the $T_{\text{c}}(P)$
dependence for the alloy begins to pull away rapidly from that of Y,
reaching a maximum suppression $\Delta T_{\text{c}}\approx $ 5 K at 81 GPa,
the highest pressure of the experiment. This strong suppression of Y's
superconductivity by dilute Tb ions points to giant Kondo pair breaking, as
has previously been observed in high pressure studies on the dilute magnetic
alloys La(Ce),\cite{lace} La(Pr),\cite{lapr} Y(Pr),\cite{fabbris1,ypr} and,
most recently, Y(Dy).\cite{fabbris1} In contrast, as seen in the inset to
Fig. 6, $T_{\text{c}}(P)$ for Y(0.5 at.\% Gd) does \textit{not} begin to
deviate markedly from that of Y metal near $59$ GPa, where Gd's volume
collapse occurs, but rather faithfully tracks Y's value of $T_{\text{c}}$ to
127 GPa, the maximum pressure of the experiment. Unlike for Tb, the magnetic
state for Gd ions in Y remains stable to 127 GPa, so that no Kondo phenomena
are expected. We thus suggest that the anomalous pressure dependences $T_{%
\text{o}}(P)$ and $\Delta T_{\text{c}}(P)$ in Tb and Y(Tb) alloy,
respectively, have their origin in Kondo physics, as does Tb's volume
collapse itself. In support of these suggestions we point out that XANES and XES experiments on Tb to extreme pressure reveal that neither a change in
valence nor a magnetic local-itinerant transition occur to a pressure of $%
\sim70$ GPa, well above the volume collapse pressure for Tb at 53 GPa.\cite%
{fabbris1}

Could perhaps an alternative explanation for the anomalously high magnetic
ordering temperatures $T_{\text{o}}$ in Tb be the effect of crystalline
electric fields? It has been argued that such fields are likely responsible
for the significant enhancement of $T_{\text{o}}$ over de Gennes scaling in
a series of \textit{R}Rh$_{4}$B$_{4}$ compounds, where\textit{\ R} is a
lanthanide.\cite{noakes,dunlap5} If the magnetic anisotropy is strong, it
has been shown\cite{noakes,dunlap5} that the crystal field enhancement can
be as large as the factor $3J_{t}/(J_{t}+1)=2.6$ for trivalent Tb where $L=3,
$ $S=3,$ and $J_{t}=6$. No crystal field effects are possible for Gd since
it carries no orbital moment ($L=0$). The lack of a sharp upturn or other
anomalies in $T_{\text{o}}$ and $\Delta T_{\text{c}}\ $in the pressure
region 60 - 127 GPa would be consistent with the certain absence of crystal
field effects in Gd. The fact that the pressure dependence of $T_{\text{o}}$
is very similar for both Gd and Tb to 53 GPa indicates that crystal field
effects in Tb, if present, are only significant for pressures above 53 GPa
where the $T_{\text{o}}(P)$ dependence becomes anomalous. In a crystal field
scenario, however, it would be difficult to understand the sharp upturn in
the suppression of superconductivity $\Delta T_{\text{c}}$ in the dilute
magnetic alloy Y(0.5 at.\% Tb) for pressures above 53 GPa. This strong
suppression of superconductivity points rather to a Kondo physics scenario
with strong Kondo pair breaking.

Further experimentation is necessary to unequivocally establish the origin
of the anomalous behavior of $T_{\text{o}}$ and $\Delta T_{\text{c}}\ $in Tb
and Y(Tb) alloy, respectively, for the pressure region above 53 GPa. Such
experiments could include an extension of the pressure range to 2 Mbar to
search for the characteristic \textquotedblleft Kondo sinkhole
behavior\textquotedblright\ in $T_{\text{c}}(P)$
observed for Y(Pr),\cite{fabbris1,ypr} La(Ce),\cite{lace} and La(Pr)\cite%
{lapr} where the $T_{\text{c}}$-suppression $\Delta T_{\text{c}}$ reaches a
maximum as the Kondo temperature $T_{\text{K}}$ passes through the
experimental temperature range, but falls off again at higher pressures
where $T_{\text{K}}$ far exceeds $T_{\text{c}}.$ Inelastic neutron or x-ray scattering studies
to extreme pressures would help establish whether crystal-field splittings
play a role in the anomalously high values of $T_{\text{o}}$\ for Tb.

In summary, measurements of the electrical resistivity of Tb metal to
extreme pressures reveal that the magnetic ordering temperature $T_{\text{o}%
} $ exhibits a highly non-monotonic pressure dependence, appearing to rise
for $P>$ 80 GPa to anomalously high values. Parallel experiments on Gd and
dilute magnetic alloys of Gd and Tb with Y suggest that under extreme
pressures Tb is transformed from a magnetically conventional lanthanide into
one with an unconventional magnetic state, perhaps involving Kondo physics,
with anomalously high values of $T_{\text{o}}$. In contrast, Gd remains a
magnetically conventional lanthanide to pressures of at least 127 GPa.%
\vspace{0.3cm}

\noindent \textbf{Acknowledgments.} The authors would like to thank T.
Matsuoka and K. Shimizu for sharing information on their high-pressure
electrical resistivity techniques used in the present study. This work was
supported by the National Science Foundation (NSF) through Grant No.
DMR-1104742 and by the Carnegie/DOE Alliance Center (CDAC) through NNSA/DOE
Grant No. DE-FC52-08NA28554. Work at Argonne National Laboratory is
supported by the US Department of Energy, Office of Science, under contract
No. DE-AC02-06CH11357.\label{biblio}

\end{document}